# Distributed Kalman filtering with minimum-time consensus algorithm

Ye Yuan, Ling Shi, Jun Liu, Zhiyong Chen, Hai-Tao Zhang and Jorge Goncalves

**Abstract**

Fueled by applications in sensor networks, these years have witnessed a surge of interest in distributed estimation and filtering. A new approach is hereby proposed for the Distributed Kalman Filter (DKF) by integrating a local covariance computation scheme. Compared to existing well-established DKF methods, the virtue of the present approach lies in accelerating the convergence of the state estimates to those of the Centralized Kalman Filter (CKF). Meanwhile, an algorithm is proposed that allows each node to compute the averaged measurement noise covariance matrix within a minimal discrete-time running steps in a distributed way. Both theoretical analysis and extensive numerical simulations are conducted to show the feasibility and superiority of the proposed method.

## I. Introduction

Distributed estimation and data fusion have many applications in a variety of fields ranging from target tracking to control of mobile networks [1]–[12]. In such scenarios, networks of sensors are employed to take measurements on various kinds of signals of interest, which are

Y. Yuan and H.-T. Zhang are with School of Automation and State Key Lab of Digital Manufacturing Equipment and Technology, Huazhong University of Science and Technology, Wuhan 430074, P.R. China.

L.Shi is with Department of Electronic and Computer Engineering, The Hong Kong University of Science and Technology, Hong Kong.

J. Liu is with Department of Applied Mathematics, University of Waterloo, Waterloo, Ontario N2L 3G1, Canada.

Z. Chen is with School of Electrical Engineering and Computer Science, The University of Newcastle, Callaghan, NSW 2308, Australia.

J. Goncalves is with Department of Engineering, University of Cambridge, CB2 1PZ, UK, and University of Luxembourg, Faculté des Sciences, de la Technologie et de la Communication, L-4362 BELVAL, Luxembourg.

H.-T. Zhang was supported by the National Natural Science Foundation of China under Grants 61322304, 61673189 and 51120155001. (*Corresponding author: H.-T. Zhang, email: zht@hust.edu.cn.*)

regarded as the state variables. Since multiple sensors are involved, it is necessary to make full use of the available measurements to yield an optimal estimate of system states.

Kalman filters [13], which are unbiased, minimum mean-squared error estimators, are commonly used to perform the required integration due to their effective estimation capabilities [14]. Such traditional integration approaches employ a single Kalman filter for the entire network to estimate system states, which is thus named as the Centralized Kalman Filter (CKF). Another typical kind of network state estimators is the decentralized Kalman filter proposed by Speyer [15]. Therein, each node consists of both a sensor and a local Kalman filter. By exchanging information with every other node in the network, each node will be able to calculate the average value of the measurements. This value is afterwards used for state estimation, which is as optimal as the CKF by using a local Kalman filter. However, high communication complexity is still a main issue of the decentralized Kalman filter. The reason lies in the all-to-all communication protocol to calculate the optimal state estimate [16].

Afterwards, this issue was alleviated with the introduction of the Distributed Kalman Filter (DKF) by Olfati-Saber [17], [18]. Therein, each node is equipped with an additional consensus filter (CF). Thus, it only needs to exchange both its measurement and its measurement noise covariance with the immediate neighbors. By this means, the CF will be able to compute the average value of the measurements asymptotically. As a result, each node obtains the optimal state estimate solely by using local information, which remarkably decreases the communication complexity. The DKF has similar asymptotic performance to the CKF, but with superior scalability and robustness to uncertainties and external noises in the network [19]. Following this line, an observer-based distributed Kalman filter was afterwards proposed in [20]. However, so far, an efficient distributed estimator with minimal convergence time to the DKF is still lacking for sensor networks.

To fulfill such an urgent yet challenging task, a novel DKF estimator is developed hereby to enable each node in the sensor network to calculate the consensus value in *finite-time*. Unlike conventional asymptotically convergence methods [21], [22], the present algorithm only uses its own past state estimates to form a Hankel matrix, with which the eventual consensus value can be calculated in a minimum number of time steps. This algorithm has improved the existing finite-time consensus algorithms in [23], [24] to a minimal time consensus [18]. The proposed method shares the same spirit of the well-known Ho-Kalman method [25]. It can substantially shorten the convergence time of the DKF to the CKF status. Meanwhile, the algorithm is robust to the uncertainties in the local measurements. Moreover, a trade-off is considered between communication cost and computation complexity. Still worth-mentioning is that, each node in the network calculates the eventual consensus value of CF and therefore guarantees a finite-time performance. Significantly, the present algorithm supports the convenient "plug-in-and-play" mode favored by modern industrial filters. Especially for the scenarios where a sensor can be naturally disabled or dropped when a new one needs to be added.

The remainder of the paper is organized as follows: Section II introduces the preliminaries of graph theory, the CF and the DKF, and then provides the problems addressed by this paper. The two DKF estimators are then developed in Section III. Numerical simulations are conducted in Section IV to show the feasibility and superiority of the proposed DKF estimators. Finally, conclusion is drawn in Section V.

Throughout the paper, the following symbols will be used. $\mathbb{R}$, $\mathbb{Z}$ and $\mathbb{N}$ denote the sets of real numbers, integers and positive integers, respectively. For a matrix $A \in \mathbb{R}^{M\times N}$, $A[i,j] \in \mathbb{R}$ denotes the $\{i,j\}^{th}$ element, $A[i,:] \in \mathbb{R}^{1\times N}$ and $A[:,j] \in \mathbb{R}^{M\times 1}$ denotes its $i^{th}$ row and $j^{th}$ column, respectively. $A[i_1 : i_2, j_1 : j_2] \in \mathbb{R}^{(i_2-i_1+1)\times(j_2-j_1+1)}$ denotes the submatrix of $A$ corresponding to the rows $i_1$ to $i_2$ and the columns $j_1$ to $j_2$. The symbol $*[i]$ represents the $i^{th}$ element of a column or row vector $*$. The symbol $e_r^{\mathsf{T}} = [0,\dots,0,1_{r^{th}},0,\dots,0] \in \mathbb{R}^{1\times N}$. $I_N$ denotes the identity matrix of dimension $N$.

## II. Preliminaries and Problem Description

### A. Graph Theory

A sensor network with $n$ nodes can be represented by a graph $\mathcal{G} = (\mathcal{V},\mathcal{E})$ of order $n$. The set of nodes, $\{\mathrm{v}_1,\mathrm{v}_2,\dots,\mathrm{v}_n\}$, is represented by $\mathcal{V}$ and the set of edges, or communication links, between each node is represented by $\mathcal{E} \subseteq \mathcal{V}\times\mathcal{V}$. A direct communication link from node $\mathrm{v}_j$ to node $\mathrm{v}_i$ is denoted by $e[j,i] = (\mathrm{v}_j,\mathrm{v}_i) \in \mathcal{E}$ and $\mathrm{v}_i$ is said to be a neighbor of $\mathrm{v}_j$ as a result of this direct connection. All neighbors of a particular node $\mathrm{v}_j$ constitute to the set $\mathcal{N}_j = \{\mathrm{v}_i \in \mathcal{V} \mid e[i,j] \in \mathcal{E}\}$. For an undirected graph, the degree of a node is the number of edges that are incident on the node. The adjacency matrix of the graph, $\widehat{A}$, refers to an $n$-by-$n$ matrix where the off-diagonal element $\widehat{A}[i,j]$ is the weight of the edge from $\mathrm{v}_j$ to $\mathrm{v}_i$. The degree matrix of the graph, $\widehat{D}$, is an $n$-by-$n$ diagonal matrix such that

$$\widehat{D}[i,j] = \begin{cases} \sum_j \widehat{A}[i,j] & \text{if } i = j; \\ 0 & \text{otherwise.} \end{cases}$$

The graph Laplacian matrix, $\widehat{L}$, is given by $\widehat{L} := \widehat{D} - \widehat{A}$.

*B. Distributed Kalman Filters*

The aim of distributed Kalman filtering is to estimate the state of a process with the following dynamics:

$$x(k+1) = Ax(k) + Bw(k), \tag{1}$$

with the $i^{th}$ sensor node having a sensing model given by:

$$z_i(k) = H_i(k)x(k) + v_i(k). \tag{2}$$

Both $w(k)$ and $v_i(k)$ are assumed to be zero-mean white Gaussian noise with covariance matrices given by: $E[w(k)w(l)^\intercal] = Q(k)\delta_{kl}$, and $E[v_i(k)v_j(l)^\intercal] = R_i(k)\delta_{kl}\delta_{ij}$, where $\delta$ is the Kronecker delta, i.e., $\delta_{kl} = 1$ for $k = l$, and $\delta_{kl} = 0$ otherwise.

Let $\widehat{x}_i(k)$ and $\overline{x}_i(k)$ be the minimum mean-squared error state estimates of the $i^{th}$ node based on the available measurement data up to time instants $k$ and $k-1$, respectively. The measurement data, noise, noise covariance matrix, and model of the CKF for time instant $k$ can be defined in terms of the individual node parameters, respectively, as $z_c := [z_1, z_2, \dots, z_n]$, $v_c := [v_1, v_2, \dots, v_n]$, $R_c := \text{diag}(R_1, R_2, \dots, R_n)$, and $H_c := [H_1; H_2; \dots; H_n]$, where $H_c$ is a column block matrix [17].

We also assume that $(A, H_c)$ is observable and $(A, H_i)$ is observable for every $i$.

The measurement of the CKF is thus given by $z_c = H_c x + v_c$, and the CKF state estimate, i.e., $\widehat{x}_c$, writes

$$\widehat{x}_c(k) = \overline{x}_c(k) + M_c(H_c^\intercal R_c^{-1} z_c - H_c^\intercal R_c^{-1} H_c \overline{x}_c(k)), \tag{3}$$

where $\overline{x}_c$ is the prior state estimate, $M_c = (P_c^{-1} + H_c^\intercal R_c^{-1} H_c)^{-1}$ and $P_c$ is the error covariance matrix of $\overline{x}_c$.

To propose the main problem, it is necessary to introduce a lemma [17] guaranteeing asymptotic performance of the DKF.

*Lemma 1: ( [17]) Suppose every node in the network obtains the average consensus value of $S^c$ and $g^c$, then by performing the following computation*

$$\begin{aligned} M_i(k) &= (P_i(k)^{-1} + S^c(k))^{-1}, \\ \widehat{x}_i(k) &= \overline{x}_i(k) + M_i(k)[g^c(k) - S^c(k)\overline{x}_i(k)], \\ P_i(k+1) &= AM_i(k)A^\intercal + BQ_i(k)B^\intercal, \\ \overline{x}_i(k+1) &= A\widehat{x}_i(k), \end{aligned} \tag{4}$$

*where $P_i(k) = nP_c(k)$, $M_i(k) = nM_c(k)$, $Q_i(k) = nQ(k)$ and*

$$\begin{aligned} S^c(k) &= \tfrac{1}{n}\textstyle\sum_{i=1}^n H_i^\intercal(k)R_i^{-1}(k)H_i(k) \\ &= \tfrac{1}{n}H_c^\intercal(k)R_c^{-1}(k)H_c(k), \end{aligned} \tag{5}$$

$$\begin{aligned} g^c(k) &= \tfrac{1}{n}\textstyle\sum_{i=1}^n H_i^\intercal(k)R_i^{-1}(k)z_i(k) \\ &= \tfrac{1}{n}H_c^\intercal(k)R_c^{-1}(k)z_c(k), \end{aligned} \tag{6}$$

*with $S^c(k)$ and $g^c(k)$ being the network-wide average inverse-covariance matrix and average measurement, respectively, then one has*

$$\lim_{k\to\infty} \widehat{x}_i(k) - \widehat{x}_c(k) = 0,$$

*where $\widehat{x}_i(k)$ and $\widehat{x}_c(k)$ are the estimates obtained by the DKF and the CKF, respectively.*

### *C. Consensus filters*

By Lemma 1, in order to obtain the same state estimates as the CKF, it is necessary for each agent to compute the average consensus values $g^c(k)$ and $S^c(k)$ by exchanging information only with its neighbors $N_i$. This can be done through consensus filters. There are three types of consensus filters that are used in the DKF algorithm in [17]: low-pass, high-pass and band-pass filters. Note that, $S_i(k)$, i.e., the estimate of $S^c(k)$ by node $i$, can be obtained as the output with $H_i^\intercal(k)R_i^{-1}(k)H_i(k)$ as the input. In this set-up, each node exchanges its measurement and covariance data with its neighbors at each time step. These data will be processed by consensus filters to obtain estimates of the average consensus values of $g^c(k)$ and $S^c(k)$. These values are then used by the local Kalman filter to calculate the state estimate at that time step.Figure showing the original DKF node setup

By Lemma 1, one has

$$\widehat{x}_i(k) = \overline{x}_i(k) + M_i(k)[g_i(k) - S_i(k)\overline{x}_i(k)]. \tag{7}$$

If $g_i$ and $S_i$ are the average consensus values, respectively, i.e., $g_i = g^c$, $S_i = S^c$, then the states estimate will asymptotically converge to that the CKF given by substituting Eqs. (5) and (6) to Eq. (7), or

$$\begin{aligned} &\widehat{x}_c(k) \\ =\ &\overline{x}_c(k) + M_c(H_c^\intercal R_c^{-1}z_c - H_c^\intercal R_c^{-1}H_c\overline{x}_c(k)) \\ =\ &\overline{x}_c(k) + nM_c(g^c(k) - S^c\overline{x}_c(k)). \end{aligned} \tag{8}$$

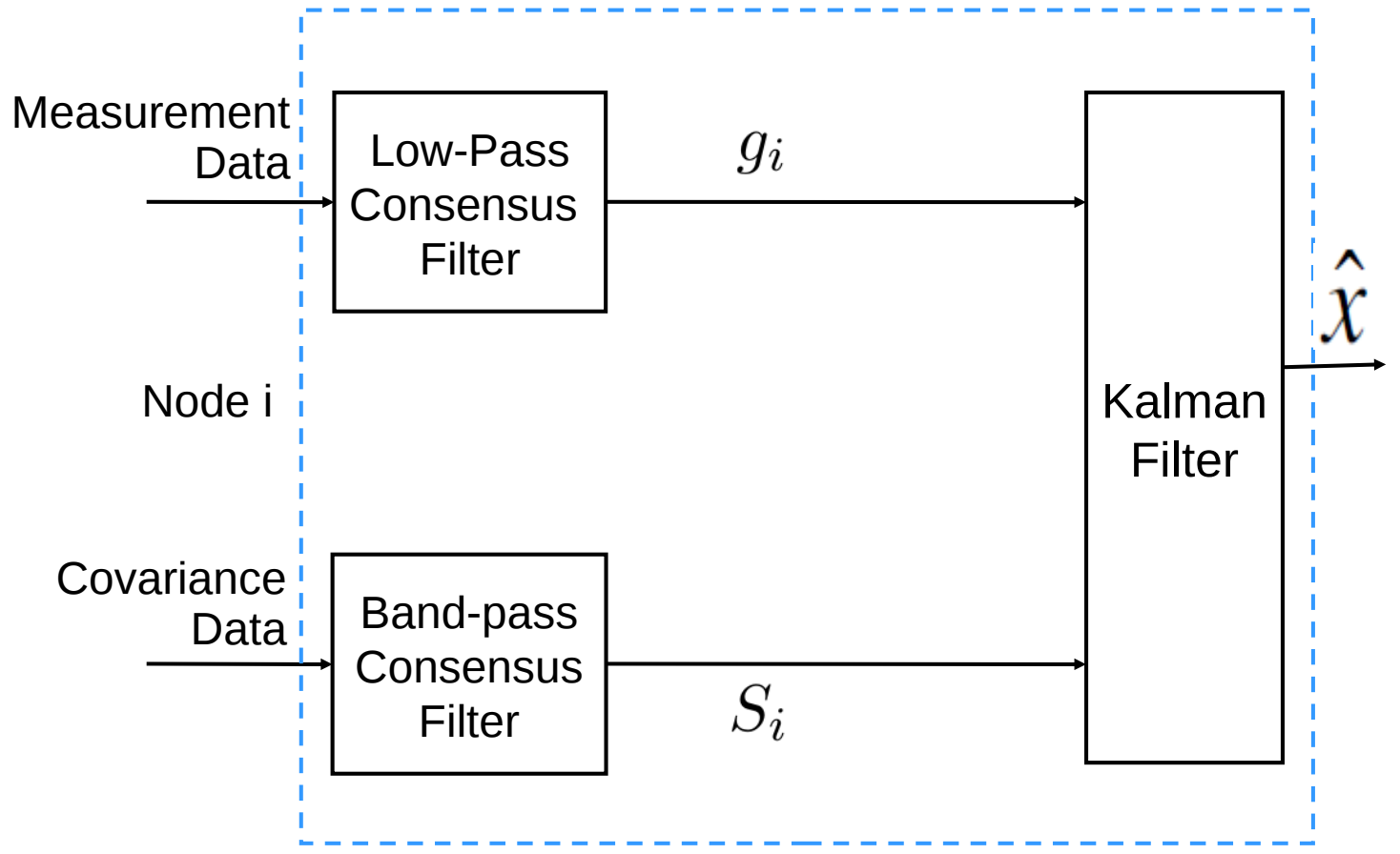

Fig. 1: Architecture for node $i$ employing the algorithm from [17]. The exchange of data from node $i$ to its neighbors $j$ is not shown in this figure.

Consider the case where consensus has not yet been obtained, the error between the state estimate $\widehat{x}_i$ by the $i^{th}$ node and state estimate $\widehat{x}_c$ of the CKF will thus be given by

$$\begin{aligned} & \widehat{x}_c - \widehat{x}_i \\ = \ & \overline{x}_c + nM_c(g^c - S^c\overline{x}_c) - \overline{x}_i - M_i(g_i - S_i\overline{x}_i) \\ = \ & \overline{x}_c - \overline{x}_i + nM_c(g^c - S^c\overline{x}_c) - M_i(g_i - S_i\overline{x}_i). \end{aligned} \tag{9}$$

## *D. Problem formulation*

The DKF algorithm [17] adopts the low-pass and band-pass CFs to compute the average consensus values $g^c$ and $S^c$ respectively. Since these CFs only obtain these values *asymptotically*, the error between the state estimates given in Eq. (9) only tends towards zero *asymptotically* as well, i.e., $\lim_{k\to\infty} \|\widehat{x}_c(k) - \widehat{x}_i(k)\|_2 = 0$. However, in industrial applications, it is unsatisfactory to asymptotically attain the convergence to the CKF estimates. Accordingly, to guarantee the convergence of the DKF estimates to the CKF estimates in finite time, rather than asymptotically, we consider two main problems this paper addresses as below,

*Problem 1: DKF estimator with finite steps of convergence:* For an $n$-node connected networked system $\mathcal{G} = (\mathcal{V}, \mathcal{E})$ governed by Eqs. (1) and (2), design a distributed estimator $u_i = \Psi z_i$ for each agent $i \in \mathcal{V}$ such that for all $t \geq M$

$$\|\widehat{x}_i(t) - \widehat{x}_c(t)\|_2 = 0$$

for arbitrary initial estimates $\widehat{x}_i(0)$ and $\widehat{x}_c(0)$. Here, $M \in \mathbb{N}$, $\widehat{x}_i(t)$ and $\widehat{x}_c(t)$ are estimates gained by the DKF and the CKF, respectively, and $\Psi$ is a compatible filter matrix to be designed.

Note that, another virtue of the above-mentioned filters lies in that it favors the requirement of "plug-in-and-play" fashion. That is, when a sensor is disabled and drops out or if a new one needs to be added– as long as the new graph is connected– the performance of underlying distributed Kalman filter should be converging to the CKF shortly. To this end, we will propose a new distributed filter development framework, where each node computes the final consensus value of CF online in a distributed way.

## III. Main Results

We consider a connected DKF network $\mathcal{G} = (\mathcal{V}, \mathcal{E})$ where the measurement noise covariance matrix for each node $i$, $R_i$, is constant. Since the measurement model, $H_i$, for each node is fixed as well, the average consensus value $S^c = \frac{1}{n}\sum_{i=1}^{n} H_i^{\intercal} R_i^{-1} H_i$ will therefore be a constant. The new framework for the DKF is proposed in Algorithm 1 (in abbr. *A1*), whose technical analysis is give below.

### A. Technical analysis of A1

Consider a discretized system of the band-pass CF in [17]

$$\begin{aligned} & S_i(k+1) \\ = \; & S_i(k) + \epsilon \Big[ \textstyle\sum_{j\in\mathcal{N}_i} (S_j(k) - S_i(k)) \\ & + \textstyle\sum_{j\in\mathcal{N}_i\cup\{i\}} (P_j(k) - S_i(k)) \Big], \\ & P_i(k+1) \\ = \; & P_i(k) + \epsilon \textstyle\sum_{j\in\mathcal{N}_i} \left[ (P_j(k) - P_i(k)) + (U_j(k) - U_i(k)) \right], \end{aligned}$$

with $U_i(k) = H_i(k)^{\intercal} R_i^{-1}(k) H_i(k)$. For arbitrary element in $S_i(k)$ and $P_i(k)$, one has

$$\begin{aligned} \begin{bmatrix} \varepsilon(k+1) \\ p(k+1) \end{bmatrix} &= \begin{bmatrix} I - \epsilon\widehat{L} - \epsilon\widehat{D} & \epsilon\widehat{A} \\ 0 & I - \epsilon\widehat{L} \end{bmatrix} \begin{bmatrix} \varepsilon(k) \\ p(k) \end{bmatrix} \\ &\quad + \begin{bmatrix} 0 \\ -\epsilon\widehat{L} \end{bmatrix} u(k) \\ &:= \mathcal{A} \begin{bmatrix} \varepsilon(k) \\ p(k) \end{bmatrix} + \mathcal{B} u(k), \\ y(k) &= e_i^{\intercal} \varepsilon(k) := \mathcal{C} \begin{bmatrix} \varepsilon(k) \\ p(k) \end{bmatrix} = S_i(k)[h, l] \end{aligned} \tag{10}$$

with

$$\begin{aligned}\varepsilon(k) &:= \begin{bmatrix} S_1(k)[h,l] & \dots & S_n(k)[h,l] \end{bmatrix}^\mathsf{T}, \\ p(k) &:= \begin{bmatrix} P_1(k)[h,l] & \dots & P_n(k)[h,l] \end{bmatrix}^\mathsf{T}, \\ u(k) &:= \begin{bmatrix} U_1(k)[h,l] & \dots & U_n(k)[h,l] \end{bmatrix}^\mathsf{T},\end{aligned}$$

and $\widehat{L}$, $\widehat{D}$, $\widehat{A}$ being the Laplacian matrix, the degree matrix and the adjacency matrix for the underlying sensor network. The sensor network is assumed to be connected so that $\widehat{L}$ matrix has a single eigenvalue at $0$ and the sampling time $\epsilon$ satisfies

$$0 < \epsilon < \frac{1}{\max\{\widehat{L}[i,i]\}}. \tag{11}$$

which guarantees consensus, as shown in [21], [26].

*Proposition 1:* Given a consensus filter in Eq. (10), all eigenvalues of $\mathcal{A}$ are within the unit disk apart from one eigenvalue at $1$.

*Proof:* This is straightforward using Gershogrin's theorem [28] and the block diagonal structure of $\mathcal{A}$. ■

Significantly, Proposition 1 guarantees the system asymptotical consensus of the constructed filter (10).

Next, we shall introduce some definitions and lemmas before giving the main results. For conciseness, we drop the subscript $i,\ j$ as the result is valid for any $i,\ j \in \mathcal{V}$.

*Definition 1:* (Minimal polynomial of a matrix pair) The minimal polynomial associated with the matrix pair $(\mathcal{A},\ \mathcal{C})$ ($\mathcal{A} \in \mathbb{R}^{2n\times 2n}$, $\mathcal{C} \in \mathbb{R}^{m\times 2n}$) denoted by $q(t) := t^{d+1} + \sum_{i=0}^{d} \alpha_i t^i$ is the monic polynomial of smallest degree $d+1$ that satisfies $\mathcal{C}q(\mathcal{A}) = 0$.

Since $q(t)$ is the minimal polynomial of the pair $(\mathcal{A},\ \mathcal{C})$, it then follows from Definition 1 that $\mathcal{C}q(\mathcal{A}) = 0$. Therefore, we obtain

$$\mathcal{C}(\mathcal{A}^{d+1} + \alpha_d\mathcal{A}^d + \ldots + \alpha_1\mathcal{A} + \alpha_0 I) = 0,$$

which immediately leads to that

$$\begin{aligned} y(d+1) &= \mathcal{C}\mathcal{A}^{d+1}x(0) \\ &= -\mathcal{C}(\alpha_d\mathcal{A}^d + \cdots + \alpha_1\mathcal{A} + \alpha_0 I)x(0) \\ &= -\alpha_d y(d) - \cdots - \alpha_1 y(1) - \alpha_0 y(0). \end{aligned}$$

One then has that the dynamics of $y(k)$ satisfies the linear regression equation:

$$\sum_{i=0}^{d+1} \alpha_i y(k+i) = 0, \quad \forall k \in \mathbb{N}, \tag{12}$$

with $\alpha_{d+1} = 1$ [27].

Denote the $z-$transform of $y(k)$ as $Y(z) := \mathbb{Z}(y(k))$. From the time-shift property of the $z-$transform, it is easy to show that

$$Y(z) = \frac{\sum_{i=1}^{d+1}\left(\alpha_i \sum_{\ell=0}^{i-1} y(\ell) z^{i-\ell}\right)}{q(z)} := \frac{H(z)}{q(z)}. \tag{13}$$

It follows from Proposition 1 that the polynomial equation $q(t) = 0$ does not possess any unstable roots except one at 1. With the polynomial $p(z) := \frac{q(z)}{z-1}$, one has that

$$p(z) = \sum_{i=0}^{d} \beta_i z^i, \tag{14}$$

for $\beta = \begin{bmatrix}\beta_0 & \dots & \beta_{d-1} & 1\end{bmatrix}^\mathsf{T}$. Then, it can be obtained that

$$\beta = \left[1 + \sum_{i=1}^{d} \alpha_i, 1 + \sum_{i=2}^{d} \alpha_i, \cdots, 1 + \alpha_d, 1\right]^\mathsf{T}. \tag{15}$$

As a result, one can calculate the consensus value $\phi$ by applying the final-value theorem and some simple algebra

$$\phi = \lim_{k\to\infty} y(k) = \lim_{z\to 1}(z-1)Y(z) = \frac{y_d^\mathsf{T}\beta}{\mathbf{1}^\mathsf{T}\beta}. \tag{16}$$

*Remark 1:* The final value of $y$, i.e., $y(\infty)$, can be computed once we know $\beta$ and the historical sequence of $y$ from $y(0)$ to $y(d)$.

Next, we propose an algorithm that obtains $\beta$ from the historical observation of $y$. Let $\overline{Y}_{0,1,\dots,2k} := \{y(1) - y(0),\ y(1), \dots, y(2k+1) - y(2k)\} (k \in \mathbb{Z})$, and consider the following Hankel matrix

$$\Gamma(\overline{Y}_{0,1,\cdots,2k}) = \begin{bmatrix} y(1)-y(0) & y(2)-y(1) & \cdots \\ y(2)-y(1) & \ddots & \\ \vdots & \vdots & y(2k+1)-y(2k) \end{bmatrix}.$$

It has been shown in [27] that when $\Gamma(\overline{Y}_{0,1,\cdots,2k})$ loses rank upon $k = K$, then $\beta$ in Eq. (16) can be calculated by

$$\Gamma(\overline{Y}_{0,1,\cdots,2K})\beta = 0. \tag{17}$$

*Proposition 2:* [27] Given a consensus filter in Eq. (10) and the corresponding minimal polynomial $q(\mathcal{A})$ in Definition 1, then the degree

$$d + 1 \leq \dim\{\mathcal{A}\} \tag{18}$$

Now we are ready to propose a new DKF algorithm that merely uses local observations to compute $S^c$ in a minimum-number of steps. From Proposition 2, we see that the minimum-number of steps for every sensor to compute the consensus value is upper-bounded by $2\dim\{\mathcal{A}\}$.

Accordingly, for an arbitrary node $i$, given the collection of successive outputs of $S_i$ from the band-pass CF,

$$\mathbb{S}_{2k}\{h,l\} = \{S_i(0)[h,l], \ldots, S_i(2k)[h,l]\} \tag{19}$$

where the indices of $S_i$ refer to the time index. One can calculate a vector of differences given by[1]

$$\begin{aligned} \overline{\mathbb{S}}_{2k}[h,l] &= \{S(1)[h,l] - S(0)[h,l], \ldots, \\ &\quad\; S(2k+1)[h,l] - S(2k)[h,l]\} \\ &= \{\overline{S}(0)[h,l], \ldots, \overline{S}(2k)[h,l]\}, \end{aligned} \tag{20}$$

at every time step $k$. This vector is then used to build up a Hankel matrix such that

$$\Gamma(\overline{\mathbb{S}}_{2k}[h,l]) = \begin{bmatrix} \overline{S}(0)[h,l] & \overline{S}(1)[h,l] & \overline{S}(2)[h,l] & \cdots \\ \overline{S}(1)[h,l] & \overline{S}(2)[h,l] & \overline{S}(3)[h,l] & \cdots \\ \overline{S}(2)[h,l] & \overline{S}(3)[h,l] & \ddots & \\ \vdots & \vdots & & \overline{S}(2k)[h,l] \end{bmatrix}. \tag{21}$$

The next step is to check whether the Hankel matrix has full rank. If not, the dimension of the Hankel matrix is increased by including the output of $S(\cdot)[h,l]$ from the next time step $k+1$, and the process is repeated afterwards. Once the Hankel matrix has lost rank or been closed to losing rank (i.e., its smallest singular value is less than a small positive threshold $\overline{\sigma}$), compute the smallest singular vector of this defective Hankel matrix, given by $K_S^{h,l} = [\beta_0^{h,l}, \ldots, \beta_{k-1}^{h,l}, 1]^\intercal$. This kernel is then used to compute the final consensus value, $S^c[h,l]$ according to the following equation:

$$S^c[h,l] = \frac{\mathbb{S}_k[h,l]^\intercal K_S^{h,l}}{\mathbf{1}^\intercal K_S^{h,l}}, \tag{22}$$

with $\mathbb{S}_k[h,l] = \begin{bmatrix} S(0)[h,l] & \ldots & S(k)[h,l] \end{bmatrix}$.

*Remark 2:* Using this approach, the consensus value $S^c$ will be obtained in minimum-time as opposed to asymptotically. Motivated by this result, Algorithm 1 is proposed which incorporates the minimum-time static consensus algorithm into the DKF.

A robust version of the proposed algorithm is presented in the Appendix.

[1]We drop index $i$ for notational simplicity.

**Algorithm 1** Distributed Kalman Filter with minimum-time consensus scheme

**Inputs:** Measurement of the state $z_i$

**Outputs:** Estimate of the state of the system $\widehat{x}$

At each time step:

*Step 1*: Update state of Low-pass Consensus Filter :

$$\begin{aligned} u_j &\leftarrow H_j^\intercal R_j^{-1} z_j, \forall j \in N_i \cup \{i\}, \\ g_i &\leftarrow g_i + \epsilon \left[ \sum_{j \in N_i} (g_j - g_i) + \sum_{j \in N_i \cup \{i\}} (u_j - g_i) \right] \end{aligned}$$

where $\epsilon$ is the sampling period.

*Step 2*: Update state of Band-pass Consensus Filter:

$$\begin{aligned} &U_j \leftarrow H_j^\intercal R_j^{-1} H_j, \forall j \in N_i \cup \{i\}, \\ &P_i \leftarrow P_i + \epsilon \sum_{j \in N_i} \left[ (P_j - P_i) + (U_j - U_i) \right], \\ &S_i \leftarrow S_i + \epsilon \left[ \sum_{j \in N_i} (S_j - S_i) + \sum_{j \in N_i \cup \{i\}} (P_j - S_i) \right]. \end{aligned} \tag{23}$$

*Step 3*: For each node $i$, compute the differences between successive values of $S_i$ (for any $h, l$): $\overline{\mathbb{S}}_{2k}[h,l] = \{S_i(1)[h,l] - S_i(0)[h,l], \ldots, S_i(2k+1)[h,l] - S_i(2k)[h,l]\}$ where $S_i(k)$ is the value of $S_i$ at the $k^{th}$ time step and form a Hankel matrix $\Gamma(\overline{\mathbb{S}}_{2k}[h,l])$. If the smallest singular value of Hankel matrix is less than a prescribed small real value $\epsilon$ at some time $k^{h,l}$, find the corresponding singular vector $K_S^{h,l} = [\beta_0^{h,l}, \ldots, \beta_{k-1}^{h,l}, 1]^\intercal$, and compute the final consensus value for $S^c[h,l]$ using eq. (22).

Then for discrete-time $k > \max_{h,l} k^{h,l}$, Steps 4 and 5 can be computed using $S^c$ instead of $S_i$, and Steps 2 and 3 itself will no longer be necessary.

*Step 4*: Estimate the state of the process using the local KF at each node:

$$M_i \leftarrow (P_i^{-1} + S_i)^{-1},$$

$$\widehat{x} \leftarrow \overline{x} + M_i (g_i - S_i \overline{x}).$$

*Step 5*: Update the state of the local KF for the next time step

$$P_i \leftarrow A M_i A^\intercal + B Q_i B^\intercal,$$

$$\overline{x} \leftarrow A \widehat{x}.$$

### B. Estimator performance analysis

A virtue of the proposed estimator saves communication cost and accelerated convergence speed to the CKF estimator. Meanwhile, the performance is preserved as that of the CKF. Once $S^c$ is calculated, communication of exchanging $S_i(k)$ is no needed. Hence the more quickly the average consensus matrix is calculated, the more communication savings we have. More precisely, the communication cost is reduced from $\mathcal{O}(n^4)$ to $\mathcal{O}(n^3)$, while the time taken for reaching the optimal performance has been reduced from $\mathcal{O}(n^2)$ to $\mathcal{O}(n)$ as well.

In terms of performance, by using Algorithm 1, it will take a much shorter time for the error between $\widehat{x}_i$ and the CKF estimates $\widehat{x}_c$ to reduce to $M_i(g^c - g_i)$, by noting that $nM_c = M_i$ when $S_i = S^c$.

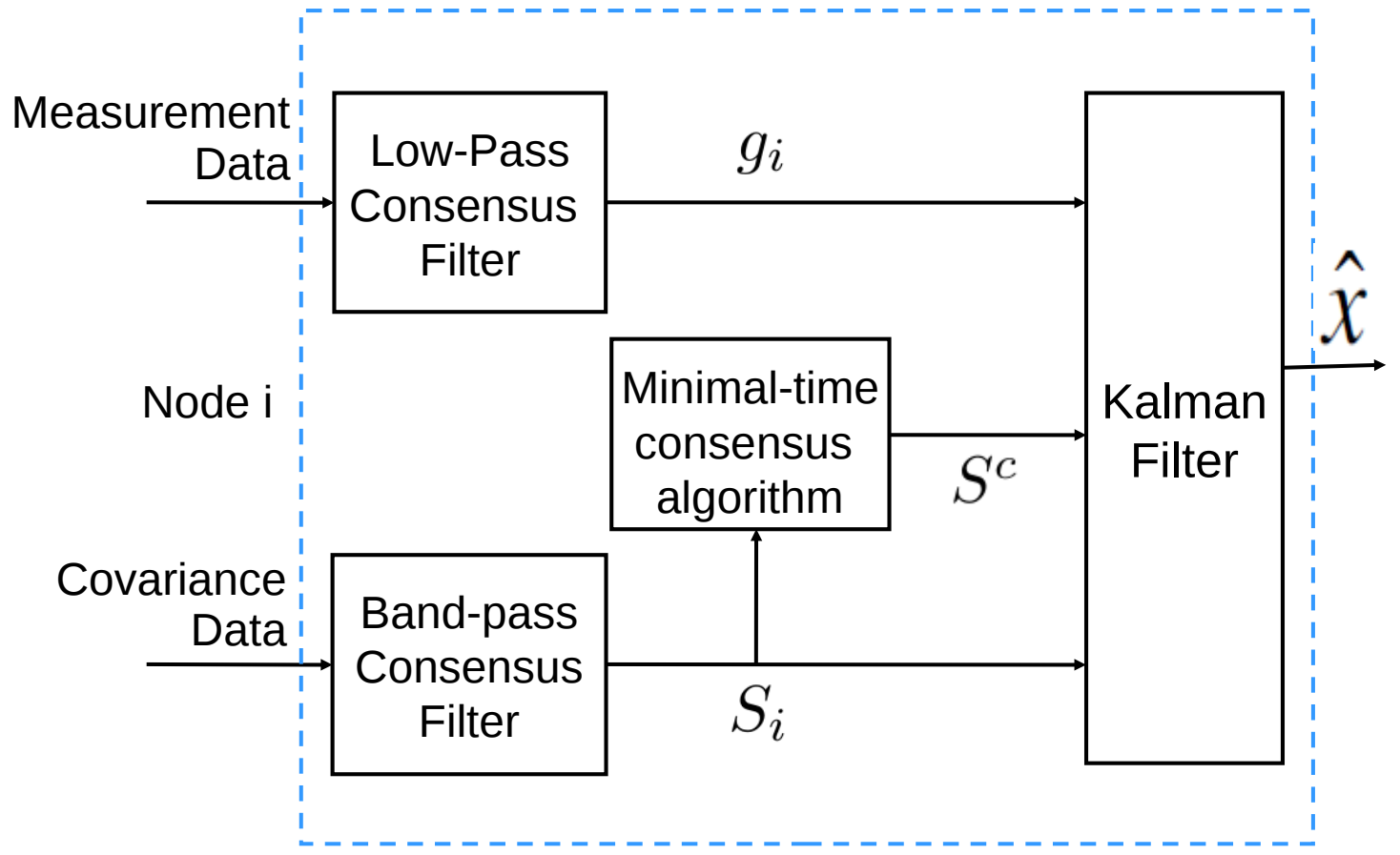

Fig. 2: Architecture for node $i$ employing Algorithm 1 [18]. In this case, we introduce a new algorithm to compute $S^c$ using $\overline{S}(k)$. The exchange of data from node $i$ to its neighbors $j$ is not shown.

A comparison is conducted between the algorithm proposed in [17] (referred to as Algorithm 0 or *A0*) , and the proposed algorithm *A1*. More precisely, for *A0*, each node $i$ exchanges its estimate on $g^c(k)$ and $S^c(k)$ with its neighbors $j$ at each time step $k$. These estimates are then transferred to the respective CFs to update the estimates, $g_i$ and $S_i$. Then a local Kalman filter is adopted to calculate the state estimate. In *A1*, the output $S_i$ of the band-pass CF is fed to the minimum-time static consensus algorithm. Once the consensus value for all elements of $S^c$ has been obtained at a time instant $t = K$, the local Kalman filter uses $S^c$ instead of $S_i$ to compute the state estimates. Therefore, using *A1*, the difference between the state estimates obtained by the DKF and the CKF will be minimized in a much shorter time.

Now, we are ready to give the main technical result as below,

*Theorem 1:* Given a connected sensor network running DKF, for any node $i$, assume that

a) the prior state estimate $\overline{x}_i = \overline{x}_c$;
b) there exists a time constant $K$ and a prescribed small positive value $\rho$ such that $\|g_i(k) - g^c(k)\|_2 \leq \rho$ for any $k > K$;

then one has

$$\|\widehat{x}_i^{A1}(k) - \widehat{x}^c(k)\|_2 \leq \|\widehat{x}_i^{A0}(k) - \widehat{x}^c(k)\|_2 \tag{24}$$

with $\widehat{x}_i^{A1}(k)$ and $\widehat{x}_i^{A0}(k)$ being the state estimates of A1 and *A0* respectively for any time $k > K$. Moreover, the equality holds if and only if $S_i(k) = S^c$.

*Proof:* The state estimate of the $i^{th}$ node is given by: $\widehat{x}_i(k) = \overline{x}_i(k) + M_i(k)[g_i(k) - S_i(k)\overline{x}_i(k)]$, and that of the CKF is given by $\widehat{x}_c(k) = \overline{x}_c(k) + nM_c(g^c(k) - S^c\overline{x}_c(k))$.

Thus the error between the state estimates of the $i^{th}$ node and that of the CKF at time $k$ will be given by:

$$\begin{aligned}
&\widehat{x}_c - \widehat{x}_i \\
=&\overline{x}_c + nM_c(g^c(k) - S^c\overline{x}_c) - \overline{x}_i - M_i(g_i(k) - S_i(k)\overline{x}_i) \\
=&\overline{x}_c - \overline{x}_i + nM_c(g^c(k) - S^c\overline{x}_c) - M_i(g_i(k) - S_i(k)\overline{x}_i) \\
=&\left((P_i(k) + S^c)^{-1} - (P_i(k) + S_i(k))^{-1}\right) g^c + \\
&(-(P_i(k) + S^c)^{-1}S^c + (P_i(k) + S^i(k))^{-1}S^i(k))\overline{x}_c \\
&+ nM_c(g^c(k) - g_i(k)).
\end{aligned}$$

Assume that $g_i(k) = g_c$ and $\overline{x}_c = \overline{x}_i$, then we can formulate the following optimization problem as: $\min_{S_i} \|\widehat{x}_c - \widehat{x}_i\|_2$. That is

$$\begin{aligned}
&\min_{S_i} \| \left((P_i(k) + S^c)^{-1} - (P_i(k) + S_i)^{-1}\right) g^c \\
&+ \left(-(P_i(k) + S^c)^{-1}S^c + (P_i(k) + S^i)^{-1}S^i\right) \overline{x}_c \\
&+nM_c(g^c(k) - g_i(k))\|_2,
\end{aligned}$$

which can be approximated as

$$\begin{aligned}
&\min_{S_i} \| \left((P_i(k) + S^c)^{-1} - (P_i(k) + S_i)^{-1}\right) g^c \\
&+ \left(-(P_i(k) + S^c)^{-1}S^c + (P_i(k) + S_i)^{-1}S_i\right) \overline{x}_c\|_2
\end{aligned} \tag{25}$$

for any $k > K$ using the triangle inequality and the assumption that $\|g^c(k) - g_i(k)\|_2$ is small.

It follows from Eq. (25) that the minimum value 0 is achieved when $S_i = S^c$ which is scheme for *A1*. Therefore, *A1* outperforms *A0*, which completes the proof. ∎

*Remark 3:* Theorem 1 shows the theoretical improvement of *A1* compared to *A0* in terms of a smaller state estimate error, as seen in Eq. (24). Empirically, the improved performance can even be gained when assumptions a) and b) are not fulfilled.

## IV. Numerical Simulations

We shall now show how the proposed new local computation can speed up the DKF. In addition to the theoretical analysis, a comparison by numerical simulations is conducted between the algorithm proposed in [17] (referred to as Algorithm 0 or *A0*) , and the proposed algorithm *A1*. The control task is to track the position of a target in two-dimensional space, i.e, $x = [x_1, x_2]^\intercal$. The target is moving in a noisy circular path with $\dot{x} = Fx + Gw$ where $w$ is a white noise process with covariance matrix $Q$ and:

$$F = \begin{bmatrix} 0 & -3 \\ 3 & 0 \end{bmatrix}, G = I_2, Q = 25I_2. \tag{26}$$

This system was then discretized with sampling time $\epsilon = 0.015s$ to yield a discrete-time system $x(k+1) = Ax(k) + Bw(k)$ with

$$A = \begin{bmatrix} 0.9990 & -0.0450 \\ 0.0450 & 0.9990 \end{bmatrix}, B = \begin{bmatrix} 0.0150 & -0.0003 \\ 0.0003 & 0.0150 \end{bmatrix}. \tag{27}$$

The sensor network used to track the target, shown in Fig. 3, consists of twenty nodes where half the nodes have sensing model matrix of $H_1$ and the other half have $H_2$:

$$H_1 = I_2,\ H_2 = \begin{bmatrix} 1 & 2 \\ 2 & 1 \end{bmatrix}. \tag{28}$$

The measurement noise covariance matrix for node $i$, $R_i$, is taken to be constant with time and is given by:

$$R_i^{-1} = \begin{bmatrix} (0.01\sqrt{i})^{-1} & 0 \\ 0 & (0.01\sqrt{i})^{-1} \end{bmatrix}. \tag{29}$$

Fig. 4 shows the evolution comparison of the variation of the first element of $S$ between *A0* and *A1*. It is observed that *A1* calculates the average consensus value $S^c$ in about 0.2s, whereas *A0* 1.6s. Table I compares the shortest, longest and average times taken across the nodes to obtain the average consensus value for all elements of $S^c$ between *A0* and *A1*. It shows that *A1* obtains $S^c$ in about 10% of the time taken by *A0*.

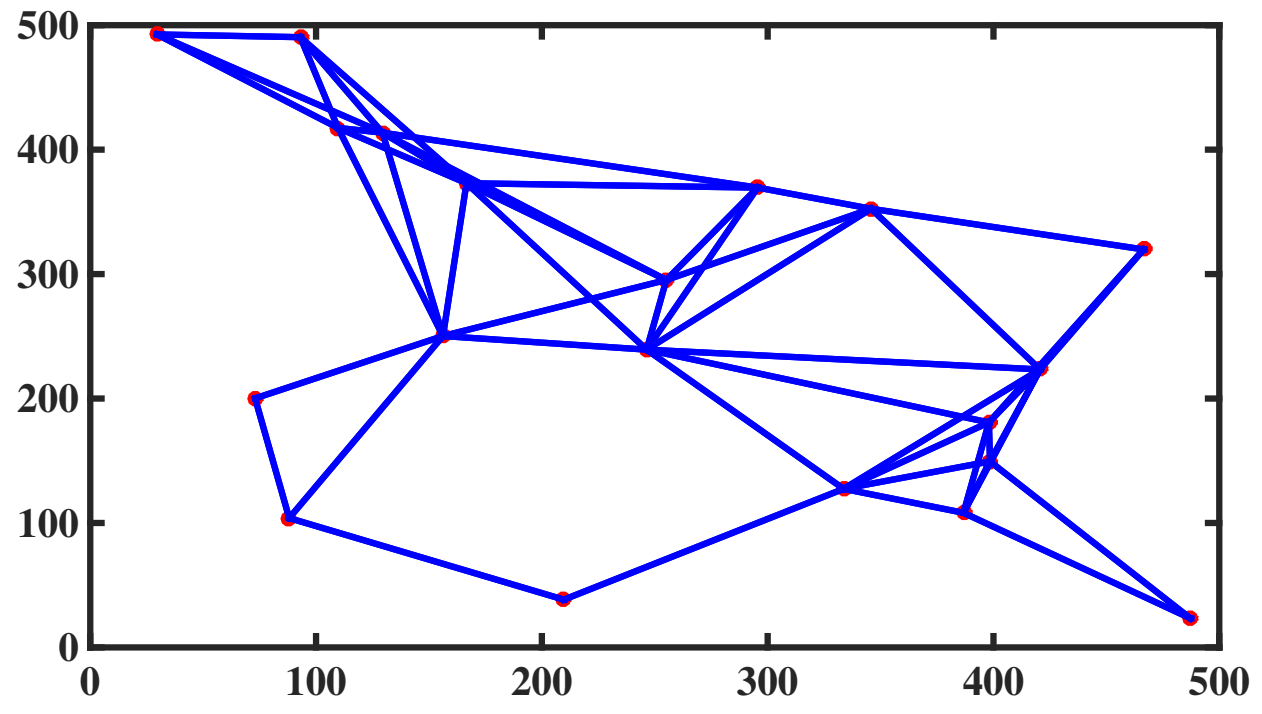

Fig. 3: A randomly generated twenty-node sensor networks. The network is undirected and connected. Red dots represent sensors while blue lines their interconnections.

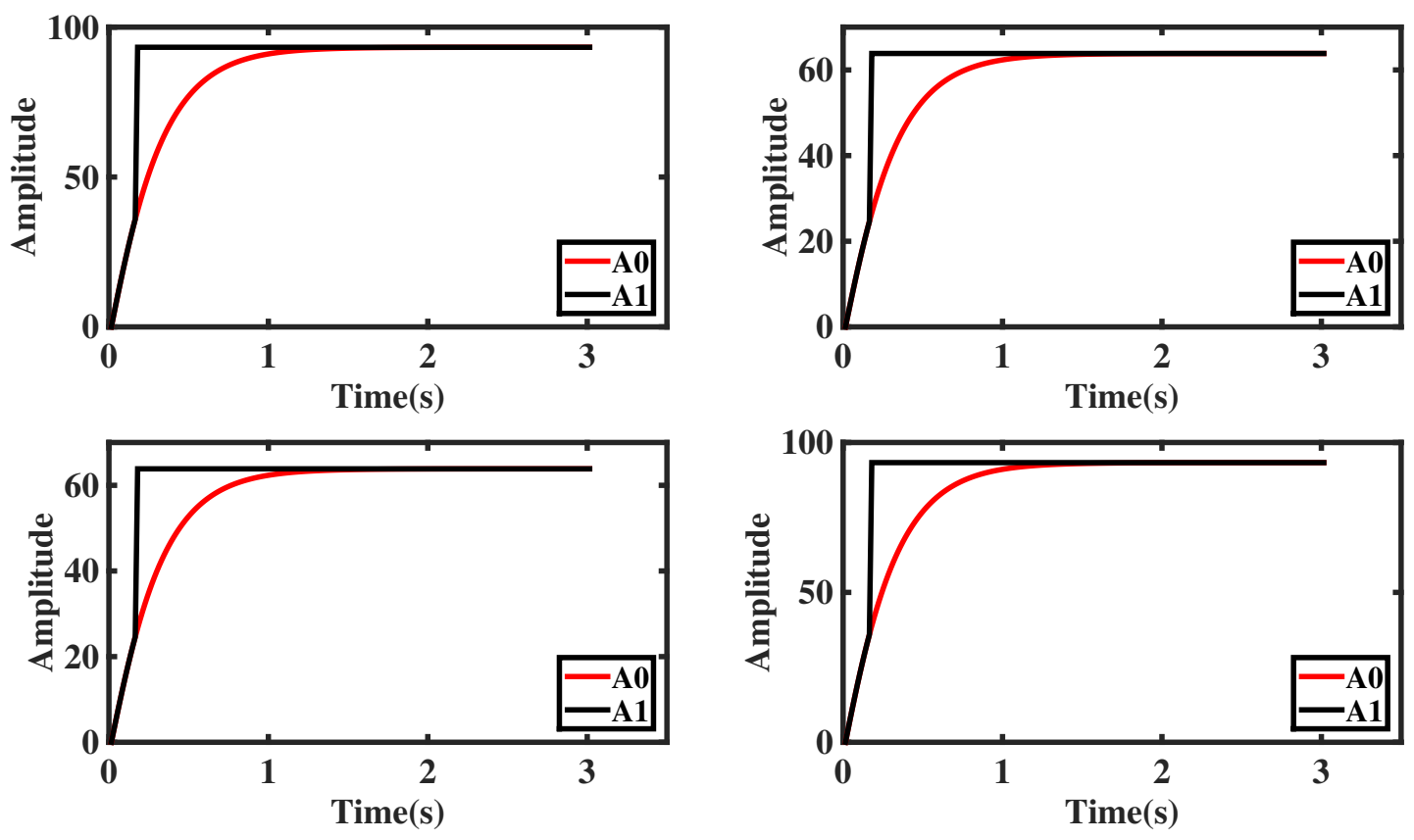

Fig. 4: The simulation results are plotted for node 19, which is randomly picked: A comparison between *A0* and *A1* on elements of $S$, clockwise from top left: $S[1,1]$, $S[1,2]$, $S[2,2]$ and $S[2,1]$. The black curves show that $S^c[i,j]$ $(i,j=1,2)$ can be calculated in a much shorter time comparing to *A0*.

To show the comparison among CFK, *A0* and *A1* more vividly, we propose a way to assess the performances in terms of state estimates. To ensure that the estimate difference are only due to the differences in values of $S$, it is assumed that all the three algorithms know the consensus value $g^c(0)$. The comparison among the three algorithms for a twenty-node network are shown in Fig. 5. Therein, in the beginning stage ($t < 0.15$s), the state estimates by both *A0* and *A1* are much larger (beyond what is shown on the graph) than that of the CKF as they have yet to obtain $S^c$. Afterwards, at $t = 0.18$s, *A1* has obtained $S^c$ according to Fig. 4, and the state estimates drastically improved until it was almost the same as that of the CKF. The slight differences from this time point onwards are due to numerical inaccuracies in the simulation. As for *A0*, the state estimates only converge to that of the CKF at about $t = 0.75$s and $t = 1$s for the states $x_1$ and $x_2$, respectively. Therefore, *A1* outperforms *A0* in terms of converge time for the state estimates to those of the CKF.

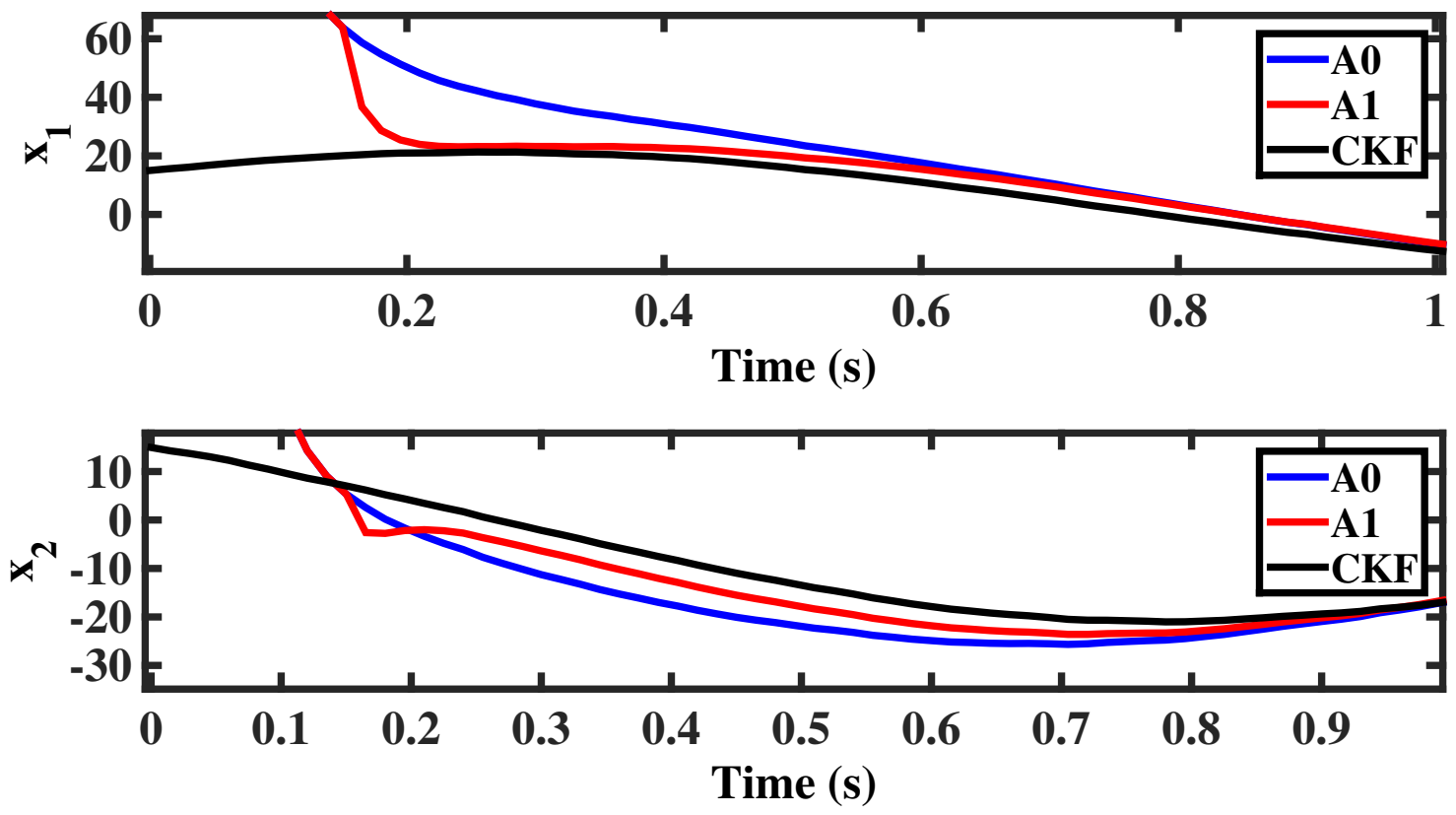

Fig. 5: Node 19 (a randomly picked node): A comparison of the state estimates by the CKF, *A0* and *A1*. The estimates by *A1* is closer to that of the CKF than *A0*, which indicates an improved performance.

| Time taken | *A0* | *A1* |
|---|---|---|
| Shortest (s) | 1.14 | 0.18 |
| Longest (s) | 1.83 | 0.21 |
| Average (s) | 1.338 | 0.1845 |

TABLE I: Times taken for *A0* and *A1* to calculate average consensus for $S$. Since different nodes have different dynamics and estimate, we look at the shortest/longest time that a node takes to know $S^c$ and the average time of the whole network to know $S^c$.

## V. Conclusion

This paper develops a DKF algorithm that enables each node in a sensor network to calculate the global average consensus matrix of measurement noise covariances without access to global information. Theoretical analyses have shown that state estimates converge to that of the CKF in a shorter time. A robust DKF is afterwards developed to overcome communication/observation noises and system uncertainties. Extensive numerical simulations are conducted to show the feasibility and superiority of the proposed DKF estimators.

## VI. Acknowledgement

We wish to thank Prof. Richard Murray for discussions. The simulation was built on Manuel Mazo Jr.'s code (TU Delft) on the realization of distributed Kalman filter (Olfati-Saber 2007 [31]) and the simulation was performed by Jerry Thia.

## APPENDIX

### A. Robust DKF algorithm, A2

In the application of *A1* to the DKF, due to the communication/observation noise and the numerical error, the Hankel matrix $\Gamma(\overline{Y}_{0,1,\cdots,2k})$ will not lose rank for any finite dimension $k$. It is necessary to propose a new robust algorithm that can incorporate both uncertainties and internal/external noises. So, we will design a robust DKF estimator to address *Problem 2*

In this end, the essential issue is to find a Hankel matrix $\Gamma(\widehat{Y}_{0,1,\cdots,2k})$ to approximate $\Gamma(\overline{Y}_{0,1,\cdots,2k})$. Since $\Gamma(\widehat{Y}_{0,1,\cdots,2k})$ has finite rank, it can be used to estimate the final consensus value [29] as follows

$$\begin{aligned}&\Gamma(\widehat{Y}_{0,1,\cdots,2k}) = \text{argmin}\|\Gamma(\overline{Y}_{0,1,\cdots,2k}) - \Gamma(\widehat{Y}_{0,1,\cdots,2k})\|,\\ &\text{s.t.: } \det\Gamma(\widehat{Y}_{0,1,\cdots,2k}) = 0, \Gamma(\widehat{Y}_{0,1,\cdots,2k}) \text{ is Hankel}\end{aligned} \tag{30}$$

where $\|\cdot\|$ can be any norm and here we pick 2-norm without loss of generality.

Accordingly, we present the following procedures (see Algorithm 2 or *A2*).

To explain how *A2* works more clearly, we seek assistance from the following lemma and then give a proposition.

*Lemma 2:* [30] Let $x \in \mathbb{R}^n$, then there exists a Hankel matrix $D \in \mathbb{R}^{n\times n}$, such that

$$Dx = x \text{ and } \|D\|_2 \leq 1.$$

*Proposition 3:* Assume that the Hankel matrix $\Gamma(\overline{Y}_{0,1,\cdots,2k})$ has full rank, then

$$\begin{aligned}&\min\|\Gamma(\overline{Y}_{0,1,\cdots,2k}) - \boldsymbol{H}(k,k)\|_2 = \underline{\sigma}(\Gamma(\overline{Y}_{0,1,\cdots,2k}))\\ &\text{s.t.: } \det \boldsymbol{H}(k,k) = 0, \boldsymbol{H}(k,k) \text{ is Hankel.}\end{aligned} \tag{31}$$

where $\boldsymbol{H}(k,k)$ can be obtained by *A2*.

*Proof:* We first define the hvec operator as a mapping from a square Hankel matrix $\mathbb{R}^{n\times n}$ to a vector $\mathbb{R}^{(2n+1)\times 1}$. For example, $\text{hvec}(\Gamma(\overline{Y}_{0,1,\cdots,2k})) = \begin{bmatrix} y_0 & y_1 & \cdots & y_{2k}\end{bmatrix}^{\mathsf{T}}$. We now propose an algorithm for computing the nearest defective Hankel matrix with respect to $\Gamma(\overline{Y}_{0,1,\cdots,2k})$.

From $A2$ , we can see that $\boldsymbol{H}(k,k)$ satisfies the constraints in optimization (30), because a). $\boldsymbol{H}(k,k)$ is Hankel by construction;

b). It is verified that $D\underline{v}(\Gamma(\overline{Y}_{0,1,\cdots,2k}) = \underline{v}(\Gamma(\overline{Y}_{0,1,\cdots,2k}))$, and hence

$$\begin{aligned} & \boldsymbol{H}(k,k)\underline{v}(\Gamma(\overline{Y}_{0,1,\cdots,2k})) \\ = & \Gamma(\overline{Y}_{0,1,\cdots,2k})\underline{v}(\Gamma(\overline{Y}_{0,1,\cdots,2k})) \\ & -\underline{\sigma}(\Gamma(\overline{Y}_{0,1,\cdots,2k}))D\underline{v}(\Gamma(\overline{Y}_{0,1,\cdots,2k})) \\ = & 0. \end{aligned}$$

As a consequence, $\boldsymbol{H}(k,k)$ does not have full rank;

c). Since $\boldsymbol{H}(k,k) - \Gamma(\overline{Y}_{0,1,\cdots,2k}) = -\underline{\sigma}(\Gamma(\overline{Y}_{0,1,\cdots,2k}))D$ and $\|D\|_2 \leq 1$, then

$$\|\boldsymbol{H}(k,k) - \Gamma(\overline{Y}_{0,1,\cdots,2k})\|_2 \leq \underline{\sigma}(\Gamma(\overline{Y}_{0,1,\cdots,2k})).$$

Therefore, $\Gamma(\widehat{Y}_{0,1,\cdots,2k}) = \boldsymbol{H}(k,k)$ solves the optimization problem (30), which completes the proof. ■

*Remark 4:* $\underline{\sigma}(\Gamma(\overline{Y}_{0,1,\cdots,2k}))$ quantifies the approximation accuracy. More specifically, if it is large (or a bad approximation), then a greater number of observations are required to increase the dimension of the Hankel matrix to gain a better approximation. Meanwhile, in A2, the Hankel matrix $\boldsymbol{H}(k,k)$ is constructed in Step 2. Besides, the c) part of the proof of Proposition 3 has shown the relationship between $D$ and $\boldsymbol{H}(k,k)$.

*Remark 5:* We shall replace Step 3 in $A1$ with the $A2$ to make the prediction robust to uncertainty.

**Algorithm 2** Decentralized robust minimum-time consensus value computation

**Data:** Successive observations of $y(i),\ i=0,1,\cdots$.

**Result:** Final consensus value: $\phi$.

*Step 1:* At each time step $k$ starting form 0, we take the singular value decomposition of $\Gamma(\overline{Y}_{0,1,\cdots,2k}) = U\Sigma V^{\mathsf{T}}$, where $\Sigma = \text{diag}\{\sigma_1,\ \sigma_2,\ldots,\sigma_{k+1}\}$ with $\sigma_1 \geq \sigma_2 \ldots \geq \sigma_{k+1} = \underline{\sigma}(\Gamma(\overline{Y}_{0,1,\cdots,2k}))$;

*Step 2:* Conduct a singular value decomposition of $\Gamma(\overline{Y}_{0,1,\cdots,2k})$ and find the smallest singular value $\underline{\sigma}(\Gamma(\overline{Y}_{0,1,\cdots,2k}))$ and corresponding singular vector $\underline{v}(\Gamma(\overline{Y}_{0,1,\cdots,2k}))$. If $\underline{\sigma}(\Gamma(\overline{Y}_{0,1,\cdots,2k})) \leq \rho$, then go to Step 3, otherwise $k = k+1$ and go to Step 1;

*Step 3:* Compute the Hankel vector

$$\text{hvec}(D) = C_x^{+} C_x^{\mathsf{T}} e_1,$$

where $C_x^{+}$ is the Moore-Penrose pseudoinverse of $C_x$, $e_1 = [1,\ 0,\ \ldots,\ 0]^{\mathsf{T}}$ has length of $2k-1$ and

$$C_x = \begin{bmatrix} v[1], & \ldots & v[k-1] & v[k] & & \\ & \ddots & & \ddots & \ddots & \\ & & v[1] & \ldots & v[k-1] & v[k] \\ v[k] & & & v[1] & \ldots & v[k-1] \\ \vdots & \ddots & & & \ddots & \vdots \\ v[2] & \ldots & v[k] & & & v[1] \end{bmatrix}.$$

*Step 4:* Let $\Gamma(\widehat{Y}_{0,1,\cdots,2k}) = \Gamma(\overline{Y}_{0,1,\cdots,2k}) - \underline{\sigma}(\Gamma(\overline{Y}_{0,1,\cdots,2k}))D$.

*Step 5:* Upon obtaining $\Gamma(\widehat{Y}_{0,1,\cdots,2k})$, we adopt Eq. (16) to compute the final consensus value.